\documentclass[final,5p,times,twocolumn]{elsarticle}
\journal{Physics Letters B}
\usepackage{slashed}
\usepackage{amsmath}
\usepackage{amssymb}
\usepackage{ulem}
\usepackage{color}
\usepackage{braket}
\renewcommand\sout{\bgroup \color{red} \ULdepth=-.5ex \ULset}

\usepackage{mathtools}
\usepackage{epsfig}
\usepackage{bm}

\newcommand{\beqn}{\begin{eqnarray}}
\newcommand{\eeqn}{\end{eqnarray}}
\newcommand{\be}{\begin{align}}
\newcommand{\ee}{\end{align}}
\newcommand{\eq}[1]{(\ref{#1})}

\newcommand{\cl}{{\mathrm{cl}}}

\newcommand{\cL}{{\cal L}}

\newcommand{\bs}{\boldsymbol}
\newcommand{\avr}[1]{{\left\langle #1 \right\rangle}}

\begin{document}

\begin{frontmatter}

\title{Anomalous Gravitational TTT Vertex, Temperature Inhomogeneity, \\ and Pressure Anisotropy}

\author[Tours,Vladivostok]{M. N. Chernodub}
\author[Lecce]{Claudio Corian\`o}
\author[Lecce]{Matteo Maria Maglio}

\address[Tours]{Institut Denis Poisson UMR 7013, Universit\'e de Tours, 37200 France}
\address[Vladivostok]{Laboratory of Physics of Living Matter, Far Eastern Federal University, Sukhanova 8, Vladivostok, 690950, Russia}
\address[Lecce]{Dipartimento di Matematica e Fisica "Ennio De Giorgi" Universit\`a del Salento and INFN Lecce, Via Arnesano, 73100 Lecce, Italy}

\begin{abstract}
The conformal anomaly in curved spacetime generates a nontrivial anomalous vertex, given by the three-point correlation function $TTT$ of the energy-momentum tensor $T^{\mu\nu}$. We show that a temperature inhomogeneity in a gas of charged massless particles generates, via the $TTT$ vertex, a pressure anisotropy with respect to the axis of the temperature variation. This very particular signature may provide an experimental access to the elusive gravitational coefficient $b$ which determines the anomaly contribution of the Weyl tensor to the trace of the energy-momentum tensor in curved spacetime. We present an estimate of the pressure anisotropy both for fermionic quasiparticles in the solid-state environment of Dirac semimetals as well as for a quark-gluon plasma in relativistic heavy-ion collisions. In both cases, the pressure anisotropy is small compared to the mean thermal pressure.
\end{abstract}

\begin{keyword}
conformal anomaly \sep 
thermodynamics \sep
curved spacetime 
\end{keyword}

\end{frontmatter}

\section{Introduction}

A physical system is defined to be scale invariant at the classical level when all the parameters of the system are dimensionless quantities. In all known physical theories, scale invariance is naturally extended to conformal invariance, and no reasonable counterexamples have been found where such an enhancement is absent \cite{Nakayama:2013is}.

In interacting theories, classical conformal invariance may break at quantum level, thus revealing the presence of a conformal anomaly~\cite{Duff:1977ay,Duff:1993wm}. A quantum anomaly, in general, regularly leads to the emergence of an associated anomalous transport law, which describes the appearance of a particular, usually unexpected in a classical theory, charge flow under the influence of specific external conditions~\cite{KLetal12}. 

While the quantum anomalies were predominantly discussed in the past in the context of particle physics~\cite{Shifman:1988zk}, nowadays  anomalies are addressed in solid-state systems as well. This may offer a reliable and systematic way for the experimental studies of their phenomenological implications~\cite{Letal14a,Letal14b,Xu15,Lvetal15,Xuetal15}. Specifically, Dirac semimetals manifest several quantum anomalies which lead to various anomaly--related transport phenomena~\cite{Karl14}. 

Dirac semimetals are three dimensional crystals whose low--energy excitations are solutions of the massless Dirac equation. Their $SO(1,3)$ Lorentz symmetry is naturally enhanced to a classical conformal $SO(2,4)$ symmetry, provided that the fermion interaction in the bulk of such materials is also translational invariant. The conformal anomaly (see \cite{Duff:1993wm} for an overview) in these materials reveals itself via the appearance of a logarithmic dependence of the photon polarization function on the renormalization scale \cite{IN12,YMetal14,JG14,RJH16,PFV18}. For such reasons topological semimetals have attracted wide research interests~\cite{NaMa16,AMV17}. 

The anomalous charge and energy flows can be described in terms of the chiral/conformal/gravitational anomaly actions, depending on the case, which play a key role in high--energy phenomenology~\cite{Coriano:2019dyc} and in heavy--ion collisions at high energy scales~\cite{KLetal12}. The axial anomaly~\cite{KK13,XKetal15,Lietal15,ZXetal16} generates -- via the chiral magnetic effect~\cite{ref:CME} -- an electric current parallel to the axis of a background magnetic field~\cite{LKetal16} which can be measured in appropriate experiments. For example, the mixed axial-gravitational anomaly~\cite{Landsteiner:2011cp} leads to a positive magneto-thermoelectric conductance for collinear temperature gradients and magnetic fields~\cite{CCetal14,Getal17}. Related axial-torsional anomalies can also be studied at experimental level, for generating an alternating electric current driven by sound waves in Weyl semimetals~\cite{Ferreiros:2018udw}. 

It has been also suggested that the conformal anomaly may generate -- via the scale magnetic effect~\cite{Chernodub:2016lbo} -- an anomalous thermoelectric current in topological semimetals, whenever a temperature gradient is present in the material~\cite{CCV18,ACV19}. The conformal anomaly produces an electric current and a current density at a boundary of a conformal system, if subjected to a background electromagnetic field~\cite{McAvity:1990we,Chu:2018ksb,Chernodub:2018ihb}. It may provide an experimental access to the beta function associated with the running of the electric charge~\cite{Chernodub:2019blw}.

Certain quantum anomalies, such as conformal and mixed axial-gravitational anomalies, may reveal themselves in curved spacetimes because they involve the energy-momentum tensor and, consequently, the metric tensor. In condensed matter systems, these ``gravitational'' anomalies may be probed in an off-equilibrium regime using the Luttinger theory of thermal transport coefficients \cite{Luttinger:1964zz,Stone:2012ud}, which was used, for example, in studies of the thermal imprints of the axial-gravitational anomaly~\cite{CCetal14,Getal17}. 

The basic idea is that the effect of a temperature gradient ${\bs \nabla} T$ -- that drives a system out of equilibrium -- can be compensated, at linear order, by a non-uniform gravitational potential $\Phi$
\beqn
\frac{1}{T} {\bs \nabla} T = - \frac{1}{c^2} {\bs \nabla}  \Phi\,,
\label{eq:Luttinger}
\eeqn
where $c$ is the speed of light. For a weak gravitational field (in the Newtonian limit), the gravitational potential $\Phi$,
\beqn
g_{00} = 1 + \frac{2 \Phi}{c^2}\,,
\label{eq:g00}
\eeqn
is related to the $g_{00}$ component of the metric, while other components of the metric tensor are unmodified. This observation is closely related to the Tolman--Ehrenfest effect~\cite{ref:TE:1,ref:TE:2}, which states that in a stationary gravitational field, the local temperature of a system at thermal equilibrium is not constant in space. The temperature depends on the spatial coordinates as
\beqn
T(x) = T_0 /\sqrt{g_{00}(x)},
\label{eq:TE}
\eeqn
where $T_0$ is a reference temperature at a selected point with $g_{00}=1$.

The Luttinger formula~\eq{eq:Luttinger} may be derived from simple thermodynamic considerations (see, for example, Ref.~\cite{ref:Rovelli}). Consider a closed system divided arbitrarily into two subsystems, 1 and 2. A thermal equilibrium happens when the total entropy, $S = S_1 + S_2$, takes its maximum, implying $d S_1 + d S_2 = 0$. If the quantity of heat $d E$ leaves the first subsystem, $d E_1 = - d E$, it always enters the second subsystem, $d E_2 = d E$, because the system is closed. For any heat exchange between the subsystems, one gets therefore $dS_1/dE_1 = dS_2/dE_2$. Given the definition of temperature, $T = dS/dE$, the last relation implies that in a thermal equilibrium, the temperatures of the two subsystems should be equal: $T_1 = T_2$. 

In a static gravitational field $\Phi$, the heat quantity $d E$ possesses an inertial mass $dm = dE/c^2$. In accordance with the equivalence between inertial and gravitational masses, the heat has a weight in a gravitational field. Therefore, the heat $dE$ leaving system 1 and entering system 2 changes its energy by performing the work against the change of the gravitational potential $\Delta \Phi = \Phi_2 - \Phi_1$ between the two subsystems: $dE_2 = d E + (\Phi_2 - \Phi_1) dm = d E_2 (1 + \Delta \Phi/c^2)$. Therefore, $T_2 = T_1 (1 + \Delta \Phi/c^2)$ and we immediately recover the Luttinger relation~\eq{eq:g00} between the gradients of temperature and the gravitational potential for closely separated nonrelativistic systems. Its relativistic generalization is shown in Eq.~\eq{eq:TE}.

In order to ensure the appropriate definition of temperature, the mentioned derivation of the Luttinger formula~\eq{eq:Luttinger} requires the sufficient proximity of the two subsystems and the weakness of the gravitation field so that $\Delta \Phi/c^2 \ll 1$. The same condition is valid in the case of a slight departure from equilibrium: the variation of the gravitational field may mimic the temperature gradient.

A crucial role in the anomalous transport is played by the quantum anomalies associated with the presence of non-vanishing 3-point functions  involving the fermions which are present in such materials. 

We recall that in a quantum field theory of chiral fermions, the nonconservation of the fermion's axial charge is generated by the $\avr{AVV}$ vertex involving the vector current $j_{V}$ and the axial current $j_{A}$. The divergence of the axial current $j_A$ is locally proportional to the product of electric and magnetic fields (represented by two ``$V$'' of the same vertex).

 The very same $\avr{AVV}$ vertex is responsible for the chiral magnetic effect: the electric current (one vector current ``$V$'') is generated in the background of a magnetic field (another ``$V$'') and of a non-zero chiral chemical potential $\mu_5$ (the time-like component of the remaining axial current ``$A$''). The chiral magnetic effect is responsible for the effect of negative magnetoresistivity, which has been experimentally observed in Weyl and Dirac semimetals.

Another example in the same theory is given by the $\avr{AAA}$ diagram, with three axial-vector currents (A), which is also responsible for the non-conservation of the axial charge in the background of an axial-vector gauge field. As is well known, axial-vector interactions, obviously, act on left- and right-handed particles with different strengths. It is however surprising that the same interaction emerges in a material. 

Indeed, the $\avr{AAA}$ vertex is responsible for a variant of the chiral magnetic effect which generates the axial current in the background of the axial magnetic field at nonzero chiral chemical potential. 
Therefore, although such chiral effects are exotic properties of the fundamental interactions in the high energy physics domain, they may readily appear in effective theories of strained Weyl semimetals. 

The $\avr{AAA}$ vertex is responsible, in particular, for the generation of a new unidirectional excitation, the chiral sound wave, for which has been recently proposed a possible experimental detection~\cite{Chernodub:2019lhw}.

In this work we are going to discuss new anomalous transport phenomena associated with the presence of another type of anomaly, the conformal/trace anomaly \cite{Duff:1993wm}. In short, the conformal anomaly implies sensitivity of certain physical phenomena on the energy scale of the interactions, in an originally scale-independent classical theory. The corresponding anomalous vertex is described by the 3-point function $\avr{TVV}$, where ``$T$'' stands for the energy-momentum tensor $T_{\mu\nu}$. 

As we are going to elaborate in more detail below, the trace of the energy-momentum tensor is a nonvanishing quantity in the classical electromagnetic background (represented by the two ``$V$'' in the diagram). For a classically conformal invariant theory the trace of the energy-momentum tensor is zero and induces an ordinary Ward identity on the TVV vertex, which is proportional to 2-point functions of vector currents (VV). 

In the quantum case this relation gets modified by the inclusion of an extra contribution given by the trace anomaly. The origin of such extra term 
can be traced back to an effective massless interaction in the form of an anomaly pole \cite{Armillis:2009pq,Giannotti:2008cv}, which in perturbation theory can be shown to be directly related to renormalization \cite{Coriano:2018zdo}. This phenomenon unifies chiral and conformal anomalies, as exemplified in the context of the supersymmetric anomaly supermultiplet in $\mathcal{N}=1$ Yang-Mills theories \cite{Coriano:2014gja}.

The anomalous 3-point function $\avr{TVV}$ diagram may also lead to anomalous transport effects. For example, the scale magnetic effect implies that in a gravitational potential (the ``$T$''), the background magnetic field (one of the $V$'s) generates an electric current (the remaining $V$) which is normal both to the gravitational and to the magnetic field. It has been suggested that such scale magnetic effect may generate Nernst thermoelectric phenomena in Dirac semimetals. 

The $\avr{TVV}$ diagram also generates the scale electric effect, which leads to the appearance of the Ohmic conductivity of the fermionic vacuum in the expanding (de Sitter) spacetime, and is indirectly related to the Schwinger effect. This phenomenon is discussed in the cosmological context in Refs.~\cite{Kobayashi:2014zza,Hayashinaka:2016qqn}. Interestingly, for theories with a positive beta function (such as QED, for example), the anomalous Ohmic conductivity is a negative quantity. 
 
In our paper we continue our investigations of the impact of the conformal anomaly on the transport and equation-of-state properties of a system of massless fermions. After a brief summary of the effects generated by the $\avr{TVV}$ diagram, we proceed with the investigation of the anomalous effects associated with the 3-point vertex $\avr{TTT}$.

\section{The Conformal anomaly in QED}
\label{sec:review}

\subsection{The flat-space case}
We consider the case of Quantum Electrodynamics (QED) with a massless Dirac fermion $\psi$ coupled to an electromagnetic field $A_\mu$. This simplest theory exhibits a variety of anomalous effects which are encountered also in more complex theories, including those that describe topological, Dirac semimetals. A discussion of the 1PI  (1--particle irreducible) conformal anomaly action in this model can be found in \cite{Armillis:2009pq} while the connection between the structure of such action, the process of renormalization, and the generation of a massless nonlocal interaction (an anomaly pole) which is the key signature of the conformal anomaly, has been discussed in \cite{Armillis:2009im} and, more recently, in \cite{Coriano:2018zdo}. Below we will discuss the structure of such anomaly poles starting from the nonlocal Riegert action 
\cite{Riegert:1984kt}, which provides an equivalent description of such exchanges, as shown for the $TVV$ and $TTT$ correlators \cite{Coriano:2018bsy,Coriano:2018bbe}. The latter (TTT) will play a key role in our current analysis. 

The Lagrangian of massless QED with a single fermion,
\beqn
\cL = - \frac{1}{4} F^{\mu\nu} F_{\mu\nu}  + {\bar \psi} i {\slashed D} \psi\,,
\label{eq:L:massless:QED}
\eeqn
involves the field strength tensor $F_{\mu\nu} = \partial_\mu A_\nu -  \partial_\nu A_\mu$ of the gauge field $A_\mu$ coupled to the Dirac four spinor $\psi$ with $D_\mu = \partial_\mu + i e A_\mu$. We consider first the model in a flat Minkowski spacetime with the metric
\beqn
\eta_{\mu\nu} = {\mathrm{diag}}(+1,-1,-1,-1),
\label{eq:eta:munu}
\eeqn
and then proceed to study the effects of a curved spacetime.

At the classical level, massless QED is characterized by a global $U(1)_L\times U(1)_R \equiv U(1)_V\times U(1)_A$ chiral symmetry. It leads to conservation of the chiral currents,
\beqn
j_{L/R}=\frac{1}{2} \int \overline{\psi}\gamma^\mu( 1 \pm \gamma_5)\psi, \qquad \partial \cdot j_{L/R}=0,
\eeqn
with the left $Q_L$ and right chiral charges $Q_R$, respectively:
\beqn
Q_L= \int d^3 x j_L^0(x, t) \qquad Q_R= \int d^3 x j_R^0(x, t).
\eeqn
At a quantum level, the ordinary gauge invariance $ U(1)_V$ is an unbroken symmetry. It leads to a zero divergency of the vector current and to the conservation of the vector (electric) charge:
\beqn
\hskip -5mm
j^\mu_V \equiv j^\mu_R+ j^\mu_L = {\bar \psi} \gamma^\mu \psi, 
\quad 
\partial \cdot j_V=0,
\label{eq:j:V}
\quad
Q = \int d^3 x j_V^0(x, t).
\label{eq:L:massless:QEDx}
\eeqn
 
The axial symmetry $U_A(1)$ is broken by quantum fluctuations signaling the existence of a quantum anomaly. The axial charge is not conserved at quantum level. In a flat spacetime, the axial current 
\beqn
j^\mu_A \equiv j^\mu_R -j^\mu_L  = {\bar \psi} \gamma^\mu \gamma^5 \psi,
\label{eq:j:A}
\eeqn
possesses a nonzero divergence in a classical electromagnetic background:
\beqn
\partial_\mu j^\mu_A = \frac{e^2}{8 \pi^2} {\widetilde F}^{\mu\nu} F_{\mu\nu} \equiv \frac{e^2}{2\pi^2} {\bs E} \cdot {\bs B},
\label{eq:d:j:A}
\eeqn
where ${\bs E}$ and ${\bs B}$ are electric and magnetic fields, respectively, and ${\widetilde F}^{\mu\nu} = (1/2) \epsilon^{\mu\nu\alpha\beta} F_{\alpha\beta}$.
A similar breaking is induced on the dilatation current 
\beqn
j_D(x) =x^\alpha T_{\alpha}^\mu \qquad  \partial\cdot  j_D=T^\mu_\mu
\eeqn
which at quantum level is promoted to the form 
\beqn
\partial\cdot  j_D=\langle T^\mu_\mu\rangle
\eeqn
and is associated to the emergence of a nonzero $\beta(e)$ function of the running coupling in the quantum theory. 
 Since this symmetry plays a central role in our analysis, we will discuss it here in more details. 
 
 The QED Lagrangian~\eq{eq:L:massless:QED} describes a conformally invariant field theory as its action $S = \int d^4 x \, \cL$ is invariant under a simultaneous rescaling of all coordinates and fields according to their canonical dimensions:
\beqn
x \to \lambda^{-1} x\,, 
\qquad 
A_\mu \to \lambda A_\mu\,, 
\qquad 
\psi \to \lambda^{3/2} \psi\,,
\label{eq:scale:transformation}
\eeqn
where $\lambda$ is a real-valued parameter. 

Scale invariance~\eq{eq:scale:transformation} is a natural outcome of the simple fact that the classical theory~\eq{eq:L:massless:QED} does not possess any characteristic mass or length scale. As a consequence, the energy-momentum tensor of the model~\eq{eq:L:massless:QED},
\beqn
T^{\mu\nu} & = & - F^{\mu\alpha} F^\nu_{\  \alpha} + \frac{1}{4} \eta^{\mu\nu} F_{\alpha\beta} F^{\alpha\beta} 
\label{eq:Tmunu:QED}\\
& & + \frac{i}{2} {\bar \psi} \left(\gamma^\mu D^\nu + \gamma^\nu D^\mu \right) \psi - \eta^{\mu\nu} {\bar \psi} i {\slashed D} \psi\,,
\nonumber
\eeqn 
is a traceless quantity on a classical level, $(T^\mu_\mu)_{\cl} \equiv 0$. 

However, at quantum level, scale invariance~\eq{eq:scale:transformation} is broken by the quantum corrections which induce a running of the electric charge $e = e(\mu)$ on the renormalization energy scale~$\mu$. In other words, the electric charge of a particle gets partially screened by quantum fluctuations. As the effectiveness of the screening depends on the distance (energy) at which the charge is probed, the effective electric charge becomes a distance- (energy-)dependent quantity. Therefore, the theory looses its conformal invariance due to effects induced by quantum fluctuations and interactions.

The loss of scale invariance in the quantum theory~\eq{eq:scale:transformation} manifests by a nonzero value of the beta-function associated with the running of the electric charge~$e$
\beqn
\beta(e) = \frac{{\mathrm d} e}{{\mathrm d} \ln \mu}\,.
\label{eq:beta:function}
\eeqn 
This dimensionless quantity parameterizes the breaking of conformal invariance of the model. \\
Due to the conformal anomaly, the expectation value of the trace of the energy-momentum tensor~\eq{eq:Tmunu:QED} acquires a nonzero expectation value~\cite{Shifman:1988zk}
\beqn
\avr{T^\alpha_{\ \alpha}(x)} = \frac{\beta(e)}{2 e}  F^{\mu\nu}(x) F_{\mu\nu}(x) \equiv \frac{\beta(e)}{2 e} \left( {\bs B}^2 - {\bs E}^2 \right), \quad
\label{eq:Tmunu:avr}
\eeqn
where ${\bs E}$ and ${\bs B}$ are electric and magnetic fields of the classical electromagnetic background.

In QED with only one flavour of Dirac fermions~\eq{eq:L:massless:QED}, the one-loop QED beta function takes the form
\beqn
\beta_{{\text{QED}}}^{\mathrm{1loop}} = \frac{e^3}{12 \pi^2}.
\label{eq:beta:QED}
\eeqn

\subsection{The anomaly in a curved background in the QED case}

\subsubsection{Anomaly action}

In a nontrivial spacetime background, the expectation value of the trace of the energy-momentum tensor acquires, in addition to the gauge contribution generated by the matter-related part~\eq{eq:Tmunu:avr} (i.e. the fermion loop), two contributions coming from the gravity side
\beqn
\avr{T^\mu_{\ \mu}} = b C^2 +b' H + c_M F_{\mu\nu} F^{\mu\nu}. \qquad
\label{eq:Tmunu:gravity}
\eeqn
In general, the coefficients $(b,b',c_M)$ are related to the number of massless scalars, fermions and spin-1  fields which may appear in the anomaly loops and are specific of a certain (classical) conformal field theory (CFT) in its Lagrangian realization. For non-Lagrangian realizations, i.e. for 
general conformal field theories, they are classified as ``conformal data'', which characterize a certain specific CFT.

The first term is given by the Weyl tensor squared 
\beqn
C^2 = C_{\mu\nu\alpha\beta} C^{\mu\nu\alpha\beta} \equiv R_{\mu\nu\alpha\beta} R^{\mu\nu\alpha\beta} - 2 R_{\mu\nu} R^{\mu\nu} + R^2/3,
\label{eq:Weyl:tensor}
\eeqn
which is expressed via the Riemann tensor $R_{\mu\nu\alpha\beta}$, the Ricci tensor $R_{\mu\nu} = R^{\alpha}_{\ \mu\alpha\nu}$, and the scalar curvature $R = R^\mu_{\ \mu}$.  The second term in Eq.~\eq{eq:Tmunu:gravity} is given by the linear combination $H = E - 2\Box R/3$, which involves the Euler (topological) density 
\beqn
E = {}^*R_{\mu\nu\alpha\beta} {}^*R^{\mu\nu\alpha\beta} \equiv R_{\mu\nu\alpha\beta} R^{\mu\nu\alpha\beta} - 4 R_{\mu\nu} R^{\mu\nu} + R^2,
\label{eq:Euler:density}
\eeqn
and the d'Alembertian differential operator $\Box \equiv \nabla^\mu \nabla_\mu$ of the scalar curvature $R$ expressed via the covariant derivative $\nabla_\mu$. Here 
${}^*R_{\mu\nu\alpha\beta} = \epsilon_{\mu\nu\mu'\nu'}R^{\mu'\nu'}_{\phantom{\mu'\nu}\alpha\beta}/2$ is a dual of the Riemann tensor.

In massless QED~\eq{eq:L:massless:QED} the coefficients $b$, $b'$ and $c$ in the trace expectation value~\eq{eq:Tmunu:gravity} are, respectively, as follows
\beqn
b = \frac{1}{320 \pi^2}, \qquad b' = - \frac{11}{11520 \pi^2},\qquad c_M = - \frac{e^2}{24 \pi^2}. \quad
\label{eq:bc}
\eeqn
The ``matter'' parameter $c_M$ is proportional to the one-loop QED beta function~\eq{eq:beta:QED}: $c_M = - \beta_{{\text{QED}}}^{\mathrm{1loop}}/(2e)$. The trace anomaly \eq{eq:Tmunu:gravity} reduces to Eq.~\eq{eq:Tmunu:avr} in a flat Minkowski spacetime~\eq{eq:eta:munu}.

The anomalous trace of the energy-momentum tensor~\eq{eq:Tmunu:gravity} is known to be generated by the nonlocal action~\cite{Armillis:2009pq,Riegert:1984kt,Mazur:2001aa,Mottola:2006ew}
\beqn 
S_{\mathrm{anom}}[g,A] & {=} & \frac{1}{8} \int d^4 x \sqrt{- g(x)}  \int d^4 y \sqrt{- g(y)} 
\label{eq:S:anom} \\
& & \hskip -20mm \cdot H(x) G^{(4)}(x,y)  \left[ 2 b C^2(y) + b' H(y) + 2 c F_{\mu\nu}(y) F^{\mu\nu}(y)\right]\!,
\nonumber
\eeqn
where $G^{(4)}(x, y)$ is the Green function the fourth-order differential operator, often called the Paneitz operator~\cite{ref:Paneitz}:
\beqn
\Delta_4 = \nabla_\mu \left( \nabla^\mu \nabla^\nu + 2 R^{\mu\nu} - \frac{2}{3} R g^{\mu\nu}\right) \nabla_\nu.
\label{eq:Delta:4}
\eeqn

A variation of the action~\eq{eq:S:anom} with respect to metric,
\beqn
\avr{T^\mu_{\ \mu}} & \equiv & - \frac{2 g_{\mu\nu}}{\sqrt{-g}} \frac{\delta S_{\mathrm{anom}}}{\delta g_{\mu\nu}},
\eeqn
provides us, indeed, with the correct expression for the one-loop trace anomaly in the curved spacetime~\eq{eq:Tmunu:gravity}. The anomaly action~\eq{eq:S:anom} is a nonlocal function of the gauge field $A_\mu$ and the metric $g_{\mu\nu}$. The nonlocality indicates that the scale anomaly is associated with a massless pole.

\subsubsection{Scale electromagnetic effects}

The anomaly action~\eq{eq:S:anom} is induced by quantum fluctuations in the background of the classical electromagnetic field $F_{\mu\nu}$ and in the presence of a background curved metric associated with an external gravitational field. It describes the response of the matter system under such off-shell external fields. 

Anomaly actions are not unique. For instance, it is possible to write down local actions containing extra degrees of freedom, which describe the breaking of the conformal symmetry with the inclusion of a Goldstone mode (a dilaton) in the low energy spectrum. Such local variants, usually derived using the Noether method~\cite{Coriano:2012dg,Coriano:2013nja} are expected to provide two complementary descriptions of the dynamical breaking of the conformal symmetry at two ends (UV/IR) of a renormalization group flow (see the discussion in \cite{Coriano:2019dyc} and in \cite{Coriano:2018zdo}).  

Since the action contains the explicit dependence on the field strength $F_{\mu\nu}$, the anomalous quantum fluctuations may carry a local electric current. The electric current, induced by the quantum fluctuations, can straightforwardly be computed using a variation of the anomaly action~\eq{eq:S:anom} with respect to the electromagnetic field $A_\mu$
\beqn
J^{\mu}(x) & = & - \frac{1}{\sqrt{-g(x)}}\frac{\delta S_{\mathrm{anom}}}{\delta A_\mu(x)}\nonumber \\
& = & - \frac{c_M}{\sqrt{-g(x)}} \frac{\partial }{\partial x^\nu } \biggl[\sqrt{- g(x)}  \, F^{\mu\nu} (x) 
\label{eq:J:anom:exact} \\
& & \cdot \int d^4 y \sqrt{- g(y)} G^{(4)}(x,y) \Bigl(E(y) - \frac{2}{3} \Box R(y)\Bigr)\biggr]\,,\nonumber
\eeqn
where the Euler topological density $E = E(x)$ is explicitly given in Eq.~\eq{eq:Euler:density}. The parameter $c$ is proportional to the QED beta function given in Eq.~\eq{eq:bc} for a single flavour. 

Equation~\eq{eq:J:anom:exact} provides us with the one--loop expression for the anomalous electric current induced by the conformal anomaly in an arbitrary classical gravitational background. Similarly to the action, the electric current~\eq{eq:J:anom:exact} is a non-local function of the metric and of the electromagnetic field~\eq{eq:S:anom}.

Working in a linear-response approach, we consider the case of a weak gravitational background. To this end it is convenient to rewrite the electromagnetic part of the anomaly action~\eq{eq:S:anom},
\beqn
S^{(1)}_{\mathrm{anom}} & = & - \frac{c_M}{6} \int d^4 x \sqrt{- g(x)} \int d^4 y \sqrt{- g(y)} \nonumber \\
& & \cdot R^{(1)}(x) \, \square^{-1}_{x,y} \, F_{\alpha\beta} (y) F^{\alpha\beta} (y)\,,
\label{eq:nonlocal}
\eeqn
in terms a small perturbation ($|h_{\mu\nu}| \ll 1$) around the flat metric,
\beqn
g_{\mu\nu} = \eta_{\mu\nu} + h_{\mu\nu}\,.
\label{eq:g:h}
\eeqn
The same expression of the anomaly action can be obtained by a perturbative analysis in QED \cite{Armillis:2009pq,Giannotti:2008cv}.

In Eq.~\eq{eq:nonlocal} the expression $\square^{-1}_{x,y}$ denotes a Green function of the flat-space d'Alembertian $\square \equiv \partial_\mu \partial^\mu$ and $R^{(1)}$ is the leading (linear in metric) double-derivative term of the Ricci scalar:
\beqn
R^{(1)} = \partial_\mu \partial_\nu h^{\mu\nu} - \eta_{\mu\nu} \square_0h^{\mu\nu}\,.
\eeqn
The indices are raised/lowered with the background metric tensor, $h^{\mu\nu} =  \eta^{\mu\alpha} \eta^{\nu\beta} h_{\alpha\beta}$. In linearized gravity the inverse metric tensor is $g^{\mu\nu} = \eta^{\mu\nu} - h^{\mu\nu}$, so that $g^{\mu\alpha} g_{\alpha\nu} = \delta^\mu_\nu + O(h^2)$.

The conformal anomaly~\eq{eq:Tmunu:avr} leads to anomalous transport effects which most straightforwardly reveal themselves in a conformally flat spacetime metric
\beqn
g_{\mu\nu}(x) = e^{2 \tau(x)} \eta_{\mu\nu}\,,
\label{eq:g:munu}
\eeqn
where $\tau(x)$ is a scalar conformal factor which vanishes at spatial infinity and $\eta_{\mu\nu}$ is the Minkowski metric tensor~\eq{eq:eta:munu}. For a weak perturbation, $|\tau| \ll 1$, one has $h_{\mu\nu} = 2 \tau \eta_{\mu\nu}$ so that $R^{(1)} = 6 \,\Box \tau$ and the leading contribution to the anomaly action~\eq{eq:nonlocal} reduces to the local expression:
\beqn
S^{(1),\mathrm{conf}}_{\mathrm{anom}} = \frac{e^2}{24 \pi^2} \int d^4 x\, \tau(x) \, F_{\alpha\beta} (x) F^{\alpha\beta} (x)\,.
\label{eq:nonlocal:tau:local}
\eeqn
Hereafter we use the parameter $c_M$ for one-flavor QED~\eq{eq:bc}.

A variation of the action~\eq{eq:nonlocal:tau:local} with respect to the electromagnetic field $A_\mu$,
\beqn
J^{\mu}(x) = - \frac{1}{\sqrt{-g(x)}} \frac{\delta S^{(1)}_{\mathrm{anom}}}{\delta A_\mu(x)}\,,
\eeqn
generates the anomalous electric current via the scale magnetic effect (SME)~\cite{Chernodub:2016lbo}
\beqn
{\bs J} & = &  \frac{2 \beta(e)}{e} {\bs \nabla} \tau(x) \times {\bs B}(x)\,.
\label{eq:SME}
\eeqn
In the presence of the electric field background $\bs E$ the conformal anomaly leads to the scale electric effect (SEE) which takes the form Ohm's law with the metric-dependent anomalous electric conductivity~$\sigma$
\beqn
{\bs J} = \sigma(x) {\bs E}(x)\,,
\qquad 
\sigma(t, {\bs x}) = - \frac{2 \beta(e)}{e} \frac{\partial \tau(t,{\bs x})}{\partial t}\,.
\label{eq:SEE}
\eeqn

The $\avr{TVV}$ vertex could also lead to the Nernst effect, which generates an electric current normal to the temperature gradient and to the axis of the background magnetic field~\cite{Chernodub:2017jcp}. The derivation follows the same steps shown above with just a few extra subtleties. Instead of the conformal factor~\eq{eq:g:munu} one uses the gravitational potential associated with the temperature gradient~\eq{eq:Luttinger} and \eq{eq:g00}. The Nernst coefficient, originating from the conformal anomaly, is proportional to the QED beta function~\eq{eq:beta:QED}, as expected.

\section{The $TTT$ vertex}
\label{sec:TTT}

\begin{figure}[!thb]
\begin{center}
\includegraphics[scale=0.75,clip=true]{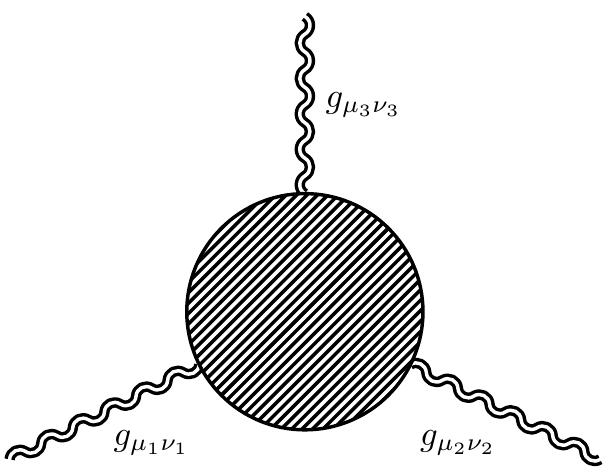} 
\end{center}
\vskip -5mm
\caption{The 1PI diagram for the TTT vertex~\eq{eq:TTT}.}
\label{fig:TTT}
\end{figure}

The anomalous contribution to the $TTT$ vertex, shown in Fig.~\ref{fig:TTT}, emerges naturally from the anomaly action~\eq{eq:S:anom} by functional differentiation. Contrary to the $TVV$ vertex, the diagram responsible for the $TTT$ vertex does not depend on the running electric charge $e$. 

As we will show below, the vertex carries information about the purely gravitational coupling $b$, which depends only on the number of fermion flavours. The value of $b$ for massless QED with a single Dirac fermion is given in Eq.~\eq{eq:bc}.

The $TTT$ vertex in momentum space can be derived in CFT by solving the conformal Ward identities using a specific procedure, starting from the transverse traceless sector of such correlator, which can be simplified by mapping the general solution \cite{Bzowski:2013sza,Bzowski:2017poo, Bzowski:2018fql} to free field theory ~\cite{Coriano:2018bsy,Coriano:2017mux}.

Building on our previous experience with the chiral magnetic and scale electromagnetic effects, we take one of the $T^{\mu\nu}$ tensors entering the $TTT$ vertex as an external probe, while the other two $T$'s are to be considered as external perturbations present in the environment. Specifically, we assume that the system is in a slightly off-equilibrium state with a constant temperature gradient in one of the directions. We will use the Luttinger identification~\eq{eq:Luttinger} to relate the variation of the $g_{00}$ component of the metric~\eq{eq:g00} with the temperature gradient~${\bs \nabla} T$. 

The $TTT$ vertex appears naturally at second order in the perturbative expansion of the effective action with respect to the metric's variations $h_{\mu\nu}$
\begin{align}
\braket{T^{\mu_1\nu_1}(x_1)}_{TTT}&= \frac{1}{8}\int dx_2\,dx_3\braket{T^{\mu_1\nu_1}(x_1)T^{\mu_2\nu_2}(x_2)T^{\mu_3\nu_3}(x_3)}\nonumber \\
&  \hskip 20mm  h_{\mu_2\nu_2}(x_2)\,h_{\mu_3\nu_3}(x_3).
\label{Kubo}
\end{align}

The anomalous part of the three-point diagram $\braket{TTT}$ in Eq.~\eq{Kubo} can be found in a straightforward way from the anomaly action~\eq{eq:S:anom}. To this end we vary the action three times respect to the metric and then take the flat spacetime limit. Ignoring the local terms, the anomaly action up to third order in the metric variation $h_{\mu\nu}$ is given by
\begin{align}
\mathcal{S}^{(3)}_{anom}[g]&=-\frac{1}{6}\int d^4xd^4x'\, R^{(1)}_x\left(\frac{1}{\square_0}\right)_{xx'} B^{(2)}_{x'}\nonumber\\
&+\frac{b'}{9}\int d^4xd^4x'\,d^4x''\,\left(\partial_\mu R^{(1)}\right)_x\left(\frac{1}{\square_0}\right)_{xx'}\label{thirdanom}\\
& \qquad \quad \cdot H^{(1),\mu\nu}_{x'}\left(\frac{1}{\square_0}\right)_{x'x''}\,\left(\partial_\nu R^{(1)}\right)_{x''},\nonumber
\end{align}
where we denoted, for brevity
\begin{align}
H^{(1),\mu\nu}_x  = \left(R^{(1)\mu\nu}-\frac{1}{3}\eta^{\mu\nu}R^{(1)}\right)_{x}, \qquad B_x^{(2)} = b\,(C^2_x)^{(2)} + b'\,E^{(2)}_x.
\nonumber 
\end{align}
The latter combination contains terms bilinear in $h_{\mu\nu}$ for, respectively, the Weyl tensor squared~\eq{eq:Weyl:tensor} and the Euler density~\eq{eq:Euler:density}.
The linear perturbations of the Riemann and Ricci tensors, as well as the scalar curvature, are taken as
\begin{align}
R^{(1)}_{\alpha\mu\beta\nu}&=\frac{1}{2}\left(\partial_\mu\partial_\beta h_{\alpha\nu}-\partial_\alpha\partial_\beta h_{\mu\nu}-\partial_\mu\partial_\nu\,h_{\alpha\beta}+\partial_\nu\partial_\alpha\,h_{\mu\beta}\right), \\
R^{(1)}_{\mu\nu}&=\frac{1}{2}\left(\partial_\mu\partial^\alpha h_{\alpha\nu}-\square_0 h_{\mu\nu}-\eta^{\alpha\beta}\partial_\mu\partial_\nu\,h_{\alpha\beta}+\partial_\nu\partial^\alpha\,h_{\mu\alpha}\right),\\
R^{(1)}&=\left(\partial^\mu\partial^\nu -\eta^{\mu\nu}\square_0\right)h_{\mu\nu}.
\end{align}

Notice that Eq.~\eq{thirdanom} may be used to calculate the three-point correlator $TTT$ provided all points are distinct from each other, $\quad x_i\ne x_j, \forall\,i\ne j$. Then all the contact terms vanish, giving
\begin{align}
& i^2\braket{\mathcal{T}\{T^{\mu_1\nu_1}(x_1)T^{\mu_2\nu_2}(x_2)T^{\mu_3\nu_3}(x_3)\}} 
\label{eq:TTT}
\\
 & =\frac{8}{\sqrt{g(x_1)}\sqrt{g(x_2)}\sqrt{g(x_3)}}\frac{\delta^3\,S[g]}{\delta g_{\mu_1\nu_1}(x_1)\delta g_{\mu_2\nu_2}(x_2)\delta g_{\mu_3\nu_3}(x_3)},
\nonumber
\end{align}
where $\mathcal{T}$ is the time-ordered product of energy-momentum tensors in the background $g_{\mu\nu}$. 

With several rearrangements, the action~\eq{thirdanom} becomes
\begin{equation}
\begin{split}
& \mathcal{S}^{(3)}_{anom}[g] = -\frac{1}{6}\int d^4xd^4x'\, \left(\partial^\mu\partial^\nu h_{\mu\nu}\right)_x\,\left(\frac{1}{\square_0}\right)_{xx'} B^{(2)}_{x'} \\
& + \int d^4x\, \left[\frac{1}{6} h(x) B^{(2)}_{x} + \frac{b'}{9} \bigl(\partial_\mu h\bigr)_x H^{(1),\mu\nu}_x \left(\partial_\nu h\right)_{x}  \right]\\
&-\frac{2b'}{9}\int d^4xd^4x'\,\left(\partial^\alpha\partial^\beta\partial_\mu h_{\alpha\beta}\right)_x\,\left(\frac{1}{\square_0}\right)_{xx'}H^{(1),\mu\nu}_{x'}\left(\partial_\nu h\right)_{x'}\\
&+\frac{b'}{9}\int d^4xd^4x'\,d^4x''\,\left(\partial^\alpha\partial^\beta\partial_\mu h_{\alpha\beta}\right)_x\,\left(\frac{1}{\square_0}\right)_{xx'}\,H^{(1),\mu\nu}_{x'}\,\\
& \qquad \cdot \left(\frac{1}{\square_0}\right)_{x'x''}\,\left(\partial^\rho\partial^\sigma\partial_\nu h_{\rho\sigma}\right)_{x''}.\label{an}
\end{split}
\end{equation}

At leading order, the anomalous contribution to the expectation value of the energy-momentum tensor~\eq{Kubo}  can be read off from the functional~\eq{an}. However, even with these simplifications, the explicit expression of the vertex~\eq{Kubo} is still very lengthy. Fortunately, for our purposes, we need only certain components of the $\avr{TTT}$ diagram.

We consider the system in a slightly off-equilibrium regime with a small temperature variation along a certain (third, in our case) direction. We use the Luttinger identification~\eq{eq:Luttinger} to relate the time-independent temperature gradient to the gradient of the gravitational potential $\Phi$. Then, the perturbation of the metric tensor is nonzero only for the $h_{00} \equiv 2 \Phi$ component, which, in addition, depends only on one spatial variable, $h_{00}(x) \equiv h_{00}(x_3)$. We find from Eqs.~\eq{an} and \eq{Kubo} that the anomalous part of the $TTT$ vertex contributes to the expectation value for the energy-momentum tensor as 
\begin{align}
\braket{T^{00}}_{TTT}&=\frac{4 b}{9} \Big[3\Big(\partial_3^2\,\Phi\Big)^2 + 4\Big(\partial_3 \Phi\Big)\,\Big(\partial_3^3 \Phi\Big) + 2 \Phi\partial_3^4\,\Phi\Big],
\label{eq:T00:pre}\\
\braket{T^{11}}_{TTT}&=\braket{T^{22}}_{TTT} = \frac{4 b}{9}\Big[2\Big(\partial_3 \Phi\Big)\,\Big(\partial_3^3 \Phi\Big) + \Phi\partial_3^4 \Phi\Big].
\label{eq:T11:pre}
\end{align}
Other components, including the energy flow $T^{0i}$ and the momentum flow $T^{ij}$, with $i,j=1,2,3$, are all equal to zero. Notice that the contribution to the pressure along the gravitational gradient is also vanishing, $P^3 \equiv T^{33} = 0$.

There are some remarkable properties of Eqs.~\eq{eq:T00:pre}--\eq{eq:T11:pre} which we need to comment upon. First, these expressions are local functions of the gravitational potential $\Phi \equiv h_{00}/2$, despite the fact that they have been derived from the non-local anomaly action. 

Second, one can readily observe that the expectation value of the energy-momentum tensor appears to involve only the anomalous coefficient $b$. This coefficient, given explicitly in Eq.~\eq{eq:bc} for the case of one-species QED, is related to the truly anomalous part of the energy-momentum tensor. Consequently, there is no topological contribution coming from the Euler density~\eq{eq:Euler:density} to the trace of the energy-momentum tensor~\eq{eq:Tmunu:gravity}.

Third, each of the nonvanishing components~\eq{eq:T00:pre}--\eq{eq:T11:pre} contains a term that depends explicitly on the gravitational potential $\Phi \equiv h_{00}/2$ itself, and not on its spatial gradient. From a condensed matter theory perspective, this property is quite surprising in view of the fact that the identification between the thermal and gravitational inhomogeneities is given in terms of their gradients~\eqref{eq:Luttinger}, and not in terms of the local temperature or the gravitational potential themselves. Notice that the $TTT$ anomalous contribution to the trace of the energy-momentum tensor,
\beqn
\avr{T^\mu_{\ \mu}}_{TTT} \equiv \avr{T^{00}}_{TTT} - \sum_{i=1}^3 \avr{T^{ii}}_{TTT} = \frac{16 b}{3} \left(\partial_3^2 \Phi \right)^2,
\eeqn
depends only on the (second) derivative of the gravitational potential.

The Tolman--Ehrenfest formula~\eq{eq:TE}, along with Eqs.~\eq{eq:g00} and \eq{eq:g:h}, allows us to derive the gravitational potential $\Phi({\bs x})$ mimicking the effect of spatially inhomogeneous temperature~$T({\bs x})$
\beqn
\Phi({\bs x}) \equiv \frac{h_{00}({\bs x})}{2}= - \frac{1}{2} \left( \frac{T^2({\bs x})}{T_0^2} - 1\right).
\label{eq:Phi:vs:T}
\eeqn
Applying a spatial gradient to both sides of Eq.~\eq{eq:Phi:vs:T} we recover, as expected, the Luttinger relation~\eq{eq:Luttinger} at leading order in the termal inhomogeneity. The ``reference'' temperature $T_0 = T({\bs x}_0)$ serves as a normalization factor: it fixes a spatial point ${\bs x}_0$ where the gravitational potential vanishes, $\Phi({\bs x}_0) = 0$.

 Let's consider a fermion gas in an off-equilibrium state with a spatially varying temperature. We assume, for simplicity, that at a point ${\bs x}$ the spatial temperature gradient takes a nonzero constant value, ${\bs \nabla} T \neq 0$, so that all the higher gradients of the temperature are vanishing, ${\bs \nabla}^n T \equiv 0$ for $n \geqslant 2$ (hereafter we promote the spatial derivative to the gradient $\partial_3 \to {\bs \nabla}$ but we always assume that the temperature varies along one fixed direction). 
 
We notice that for a linearly varying temperature, the contribution of the $TTT$ anomaly to the pressure~\eq{eq:T11:pre} vanishes, $\delta P^{i} \equiv \avr{T^{ii}}_{TTT} = 0$ in all directions $i=1,2,3$. The leading-order contribution of the $TTT$ vertex to the energy density~\eq{eq:T00:pre} is, however, nonzero. It is proportional to the fourth power of the temperature gradient,
\beqn
\delta E \equiv \avr{T^{00}}_{TTT}  = \frac{4 b \hbar c}{3} \left( \frac{{\bs \nabla} T}{T} \right)^4 \equiv \frac{\hbar c }{240 \pi^2}  \left( \frac{{\bs\nabla} T}{T} \right)^4,
\label{eq:deltaE}
\eeqn
where the last equation is given for one fermion flavour~\eq{eq:bc}. We have also restored missing powers of the Planck constant $\hbar$ and the velocity $c$ of the massless relativistic particle. 

The $TTT$ vertex of the conformal anomaly action leads to a qualitatively new effect, as it makes the pressure anisotropic with respect to the axis of the temperature variation. For this purpose it is convenient to introduce the pressure asymmetry, which characterizes the difference between the pressures along the axis of the temperature gradient and the normal respect to the same axis
\begin{align}
\delta P = P_\| - P_\perp, 
\qquad 
P_\|  = \avr{T^{33}}, 
\quad
P_\perp = \frac{\avr{T^{11}} + \avr{T^{22}}}{2}.
\label{eq:deltaP:definition}
\end{align}
According to Eq.~\eq{eq:T11:pre}, the temperature inhomogeneities may give a nonzero anomalous contribution to the pressure asymmetry~\eq{eq:deltaP:definition} of the interacting gas, provided the temperature inhomogeneities are beyond the linear regime. Assuming that that the second-order derivatives of temperature are non-zero, ${\bs \nabla}^2 T \neq 0$, we get, at leading order in the thermal gradient
\beqn
\delta P = \frac{16 b}{3} \hbar c \left( \frac{{\bs \nabla} T}{T} \right)^2  \left( \frac{{\bs \nabla}^2 T}{T} \right) \equiv \frac{\hbar c }{60 \pi^2} \left( \frac{{\bs \nabla} T}{T} \right)^2  \left( \frac{{\bs \nabla}^2 T}{T} \right),
\label{eq:deltaP}
\eeqn
where the last result is given for one fermion flavour~\eq{eq:bc}. 

In order to estimate the magnitude of the contribution of the $TTT$ anomaly in the energy density~\eq{eq:deltaE} and the pressure asymmetry~\eq{eq:deltaE} of the interacting fermion gas, it is worth comparing these quantities respectively, to the thermal energy density and to the pressure, $E_{\mathrm{th}} = 3 P_{\mathrm{th}} = 7 \pi^2 T^4/60$
\beqn
\frac{\delta E}{E_{\mathrm{th}}} = \frac{1}{28}  \left( \frac{\hbar c {\bs\nabla} T}{\pi T^2} \right)^4,
\qquad
\frac{\delta P}{P_{\mathrm{th}}} = \frac{3}{7}   \left( \frac{{\bs \nabla} T}{\pi T^2} \right)^2  \left( \frac{{\bs \nabla}^2 T}{\pi^2 T^3} \right).
\label{eq:deltaE:Eth}
\eeqn

In a solid-state environment, the conformal anomaly may be studied in the context of the Dirac semimetals, where the massless particles could be realized, for example, as fermionic quasiparticles at low energies. These excitations propagate with a Fermi velocity which is much smaller than the speed of light, $v_F \approx c/300$ (for semimetals, one should therefore replace $c \to v_F$ in all the appropriate places). To estimate the effectiveness of the $TTT$ anomaly~\eq{eq:deltaE:Eth}, one may take the pressure gradient of one Kelvin per millimeter, $\nabla T = 1. \mathrm{\,K/mm}$, of a Dirac semimetal kept at the ambient temperature $T = 10 \mathrm{\,K}$. Then we get from Eq.~\eq{eq:deltaE:Eth} an unobservable tiny energy contribution: $\delta E/ E_{\mathrm{th}} \sim  10^{-19}$. While this number may be higher for larger temperature gradients, even taking a 1 K  temperature difference at the ends of a short 1.\,$\mu$m-long rod), the contribution of the conformal anomaly is still a rather small quantity: $\delta E/ E_{\mathrm{th}} \sim 10^{-7}$. We expect the same order of magnitude, at best, for the relative pressure anisotropy~\eq{eq:deltaE:Eth}.

In the particle-physics context, the effect may take place in an expanding fireball of  quark-gluon plasma which is created in a heavy-ion collision, in experiments at the LHC at CERN or at RHIC at BNL~\cite{Busza:2018rrf}. A typical initial temperature of the fireball is a few critical temperatures $T_c \simeq 150 \mathrm{\,MeV}$. Taking $T = 2 T_c \simeq 300 \mathrm{\,MeV}$ and assuming a moderate temperature gradient, $\nabla T = 0.1\, T/\mathrm{\,fm} \simeq 30 \mathrm{\,MeV/fm} \simeq 6\times 10^3 \mathrm{\,MeV}^2$,  one gets from Eq.~\eq{eq:deltaE:Eth} very small values for the energy variation and the pressure anisotropy: $\delta E/ E_{\mathrm{th}} \sim \delta P/ P_{\mathrm{th}} \sim 10^{-8}$.

\begin{figure}[!thb]
\begin{center}
\includegraphics[scale=0.3,clip=true]{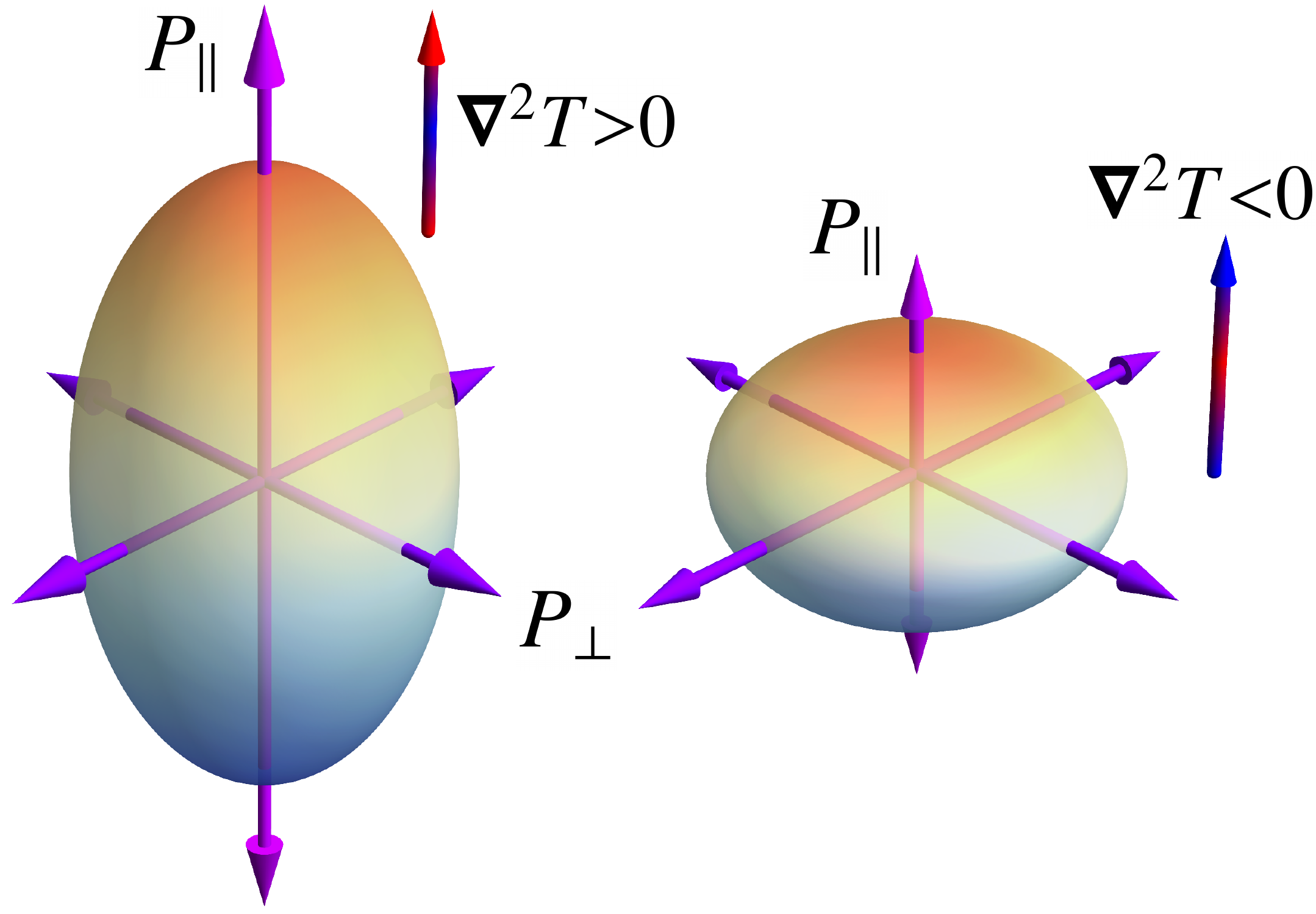} 
\end{center}
\vskip -5mm
\caption{Illustration of the anomaly-induced asymmetry~\eq{eq:deltaP} between the pressure components in the transverse plane ($P_\perp$) and in the longitudinal directions ($P_\|$) along the temperature inhomogeneities $T = T(x_\|)$ (the dimensions illustrate the strength of the components of the pressure and not the geometry of the system).}
\label{fig:illustration}
\end{figure}

The pressure asymmetry \eq{eq:deltaP} depends on the spatial concavity/convexity of the local temperature. The pressure along the axis of the temperature variation
is larger (smaller) than the pressure in the transverse directions provided ${\bs \nabla}^2 T {>} 0$ (${\bs \nabla}^2 T\, {<} 0$), as illustrated in Fig.~\ref{fig:illustration} Despite the absence of a particularly small pre-factor in Eq.~\eq{eq:deltaE:Eth}, the relative pressure asymmetry is expected to be a small number due to the high power of the relative temperature gradients due to inhomogeneities.

One may also consider the possibility that the effect may become more relevant in the astrophysical domain, in the early Universe, where the expanding gas of hot relativistic particles may experience large temperature gradients due to inhomogeneities.

\section{Conclusions}

In this work we have shown that the conformal anomaly leads to a qualitatively new effect. Specifically, the temperature inhomogeneity in a gas of interacting massless particles produces a pressure anisotropy with respect to the axis of the temperature variation. The effect originates from a purely gravitational part of the anomalous vertex given by the three-point correlator $TTT$ of the energy-momentum tensor. This phenomenon may appear in several and rather different physical scenarios: 1) in the solid-state environment of Dirac semimetals, 2) in the expanding fireballs of the quark-gluon plasma and, perhaps, 3) in the astrophysical relativistic plasmas generated in the early Universe. Although the relative pressure asymmetry is parametrically very small in the environments which have been studied so far at experimental level, the effect, nevertheless, may be used to probe the elusive gravitational coefficient $b$ which determines the anomalous contribution of the Weyl tensor to the trace of the energy-momentum tensor.

\section*{Acknowledgments}
We thank Stefania D'Agostino (IIT-Lecce), Yago Ferreiros (KTH-Stockholm), Karl Landsteiner (IFT-Madrid), Mar\'ia Vozmediano (ICMM-Madrid) for illuminating discussions. This work was partially supported by Grant 3.6261.2017/8.9 of the Ministry of Science and Higher Education of Russia. The work of C.C. and M.M.M. is partially supported by the INFN Iniziativa Specifica QFT-HEP.

\end{document}